\documentclass[apj]{emulateapj}
\usepackage{graphicx}
\usepackage{natbib}
\usepackage{amssymb}

\begin{document}

\title{Universal Relations for Innermost Stable Circular Orbits around Rapidly Rotating Neutron Stars }
\author{Shun-Sun Luk and Lap-Ming Lin }
\affiliation{Department of Physics, The Chinese University of Hong Kong, Hong Kong, China}
\email{Correspondence to: lmlin@phy.cuhk.edu.hk} 
\date{\today}
\begin{abstract}

We study the innermost stable circular orbit (ISCO) of a test particle around rapidly rotating neutron stars. Based on 12 different nuclear-matter equations of state (EOS), we find numerically two approximately EOS-insensitive universal relations that connect the radius and orbital frequency of the ISCO to the spin frequency $f$ and mass $M$ of rotating neutron stars. The relations are EOS-insensitive to about the 2\% level for a large range of $Mf$. 
We also find that the universal relation for the ISCO radius agrees with the corresponding relation for the Kerr black hole to within 6\% up to $Mf=5000 M_\odot {\rm Hz}$. Our relations can be applied to accreting neutron stars in low-mass X-ray binaries. Using the spin frequency $f=414$ Hz and the highest kilohertz quasi-periodic oscillations (kHz QPOs) at 1220 Hz observed in the system 
4U 0614+09, we determine the mass of the neutron star to be $2.0 M_\odot$. 
Our conclusion only makes a minimal assumption that the highest kHz QPO frequency is the ISCO frequency, bypassing the assumption of slow rotation and the uncertainty related to the dimensionless spin parameter, which are commonly required in the literature.

\end{abstract}

\maketitle

%%%%%%%%%%%%%%%%%%%%%%%%%%%%%%
\section{Introduction}
\label{sec:intro}
%%%%%%%%%%%%%%%%%%%%%%%%%%%%%%%

Though they were first discovered 50 years ago \citep{Hewish:1968_709}, neutron stars are still not well understood compared to other more common stellar objects like the sun. 
A major challenge to our understanding of neutron stars is identifying the properties of the poorly understood high dense nuclear matter that exists in their cores. This uncertainty is reflected
by the fact that a large number of nuclear-matter equation of state (EOS) models with various different predicted properties of neutron stars have been proposed in the past 50 years or so. 
The observed properties of neutron stars, such as their masses and radii, can thus be used to
put constraints on the theoretical EOS models \citep[see, e.g.,][]{Lattimer:2012_485}.

As the structure and properties (both static and dynamical) of a neutron star in general depend sensitively on the underlying EOS model, it is thus quite interesting that various approximately 
EOS-insensitive universal relations for neutron stars have indeed been found. These relations generally connect different physical quantities of neutron stars, and they are said to be universal in the sense that they are insensitive to EOS models to the $O(1\%)$ level (see \citet{Yagi:2017_1} for a review).

For instance, \citet{Lau:2010_1234} found a pair of universal relations to connect the frequency and damping rate of the quadrupolar $f$-mode to the mass $M$ and moment of inertia $I$ of nonrotating
neutron stars. It is also shown in \citet{Lau:2010_1234} that the values of $M$, $I$, and the stellar radius $R$ of a neutron star can be inferred accurately from the $f$-mode gravitational wave signals (if detected) emitted from the star.   
More recently, the so-called I-Love-Q relations discovered by \citet{Yagi:2013_023009,Yagi:2013_365} have gained a lot of interest. These relations connect the moment of inertia $I$, the tidal deformability (also called the Love number), and the spin-induced quadrupole moment $Q$ of slowly rotating neutron stars. 
In particular, the {\it I}$-${\it Q} relation has been extended to include rapid rotation. 
It was initially found by \citet{Doneva:2014_L6} that the {\it I}$-${\it Q} relation is broken and becomes more EOS-dependent when considering rapidly rotating stars with a fixed rotation frequency $f$. 
However, it was then found by \citet{Pappas:2014_121101} that the {\it I}$-${\it Q} relation remains approximately EOS-insensitive if, instead of the dimensional quantity $f$, one uses the dimensionless spin parameter $j$ to characterize rotation. This conclusion is extended by \cite{Chakrabarti:2014_201102} to include other dimensionless parameters such as $Mf$ or $Rf$. 
In effect, there still exists a universal {\it I}$-${\it Q} relation for each value of $j$ or $Mf$ for rapidly rotating neutron stars.

It is noted that previously known universal relations connect the intrinsic stellar properties of a neutron star itself (e.g., its mass) and the dynamical response of the star (e.g., its tidal deformability) to external perturbations.
In this work, we propose a pair of universal relations that connect the radius and orbital frequency of the innermost stable circular orbit (ISCO) to the mass and spin frequency of 
rapidly rotating neutron stars. 
The ISCO is an important prediction of general relativity concerning the strong field spacetime around a compact stellar object\footnote{It should be mentioned that an ISCO can also appear
in Newtonian gravity around highly oblate objects, such as rapidly rotating low-mass strange stars
\citep{Zdunik:2001_087501,Kluzniak:2013_2825}.}. 
It is linked to the geodesic motion of a test particle that orbits close to the compact object. 
As we shall discuss below, the ISCO universal relations are practical from an observational perspective since the ISCO may be closely related to the kilohertz quasi-periodic oscillations (kHz QPOs) well observed from neutron stars in low-mass X-ray binaries (LMXBs).

The plan of this paper is as follows. Section~\ref{sec:method} presents our numerical methods and 
chosen EOS models used in this work. The ISCO around rotating compact objects is also reviewed 
briefly.
Our main numerical results and the proposed ISCO universal relations are discussed in Section~\ref{sec:results}. In Section~\ref{sec:analy}, we compare the relation for the ISCO 
radius to the corresponding relation for the Kerr black hole and also study its connection to 
other known universal relations for rotating neutron stars. 
Finally, Section~\ref{sec:discuss} discusses 
the astrophysical relevance of the ISCO universal relations. Unless otherwise noted, we use geometric units where $G=c=1$.

%%%%%%%%%%%%%%%%%%%%%%%%%%%%
\section{Rotating neutron stars and ISCO in general relativity}
\label{sec:method}
%%%%%%%%%%%%%%%%%%%%%%%%%%

The computation of rotating stellar models in general relativity is a nontrivial task. Nevertheless, a few public codes using different formulations and numerical methods to construct rapidly rotating neutron stars in general relativity are readily available (see \citet{Paschalidis:2017_7} for a review). In this work, we use the numerical code {\tt rotstar} from the publicly available C++ LORENE library\footnote{http://www.lorene.obspm.fr} that solves the 
Einstein equations in a stationary and axisymmetric spacetime assuming a perfect-fluid matter source
using a multidomain spectral method \citep{Bonazzola:1993_421,Bonazzola:1998_104020}.    

One needs to provide an EOS for nuclear matter to compute neutron star models. While the EOS 
in the high-density core of neutron stars is still not well understood, the observations
of neutron stars with masses $M \approx 2 M_\odot$ \citep{Demorest:2010_1081,Antoniadis:2013_6131} have already ruled out many soft EOS models in general relativity. 
Note, however, that some EOS models that are ruled out by this $2 M_\odot$ constraint would be revived in other theories of gravity \citep[see, e.g.,][]{Pani:2012_084020,Sham:2012_064015}. 
In this study, we assume general relativity is the correct theory of gravity and employ 12 different nuclear-matter EOS models based on various theoretical approaches, including nuclear many-body theory, nuclear energy density functional theory, and Skyrme mean-field models. 
The first nine EOSs are APR \citep{Akmal:1998_1804}, AU (the AV14+UVII model in \citet{Wiringa:1988_1010}), BSk20, BSk21 \citep{Potekhin:2013_A48}, GM1 \citep{Glendenning:1991_2414}, SKa, SKI2 \citep{Gulminelli:2015_055803}, 
SLy4 \citep{Douchin:2000_107}, and UU (the UV14+UVII model in \citet{Wiringa:1988_1010}). The remaining three EOSs are the three representative models (soft, intermediate, and stiff) presented in \citet{Hebeler:2013_11}, which are based on nuclear interactions derived from chiral 
effective field theory combined with observational constraints. All of our chosen EOS models can support a nonrotating neutron star with a maximum mass larger than $2 M_\odot$ in order to be consistent with the $2 M_\odot$ observational constraint \citep{Demorest:2010_1081,Antoniadis:2013_6131}.

In this work, we are interested in the ISCO of test particles in a prograde orbit on the equatorial plane around rotating neutron stars 
\citep{Miller:1998_793,Zdunik:2000_612,Bhattacharyya:2011_3247,Torok:2014_L5,Cipolletta:2017_024046}.
The spacetime outside a nonrotating neutron star is described by the Schwarzschild metric and it is well known that the circumferential radius of the ISCO is given by $R_{\rm ISCO} = 6 M$ in this case. In contrast to coordinate radius, the circumferential radius is more meaningful physically and hence we shall adopt this as our definitions for both the ISCO and stellar radii in this work. 
While the ISCO radius for a nonrotating star depends only on $M$, the situation for rotating stars 
is more complicated. To first order in the dimensionless spin parameter $j\equiv J/M^2$, where $J$ is the angular momentum, the spacetime of a slowly rotating star is determined uniquely by $M$ and $j$ and the ISCO radius is given by \citep[e.g.,][]{Miller:1998_793}
\begin{equation}
R_{\rm ISCO} = 6 M  \left[ 1 - j \left({2\over 3} \right)^{3/2} \right] . 
\label{eq:Risco_slow}
\end{equation} 
As we shall see below, the exact numerical results of $R_{\rm ISCO}$ for rotating neutron stars
start to deviate from the prediction of Equation~(\ref{eq:Risco_slow}) when $j\gtrsim 0.1$. 
For a rapidly rotating neutron star, nonspherical deformation of the star is significant and 
the contributions from higher-order multipole moments of the star to the exterior vacuum spacetime 
become more important (see Section~\ref{sec:other_relations}).

For comparison, the ISCO radius around a Kerr black hole depends only on the two parameters $M$ and $j$, and it can be obtained analytically. The Boyer$-$Lindquist radial coordinate $r_{+}$ of the ISCO for a prograde orbit around a Kerr black hole is given by \citep{Bardeen:1972_347} 
\begin{equation}
r_{+}  = M \left[ 3 + Z_2 - \sqrt{ (3 - Z_1)(3+Z_1 + 2Z_2 ) } \right] , 
\end{equation} 
where $Z_1$ and $Z_2$ are defined by 
\begin{equation}
Z_1 = 1 + (1 - j^2)^{1/3} \left[ (1+j)^{1/3} + (1-j)^{1/3} \right] , 
\end{equation}
\begin{equation}
Z_2 = \sqrt{3 j^2 + Z_1^2 } .
\end{equation}
The more physical circumferential radius $R_{\rm ISCO}^{\rm Kerr}$ of the ISCO is related to 
$r_{+}$ by 
\begin{equation}
 R_{\rm ISCO}^{\rm Kerr} = \sqrt{ r_{+}^2 + a^2 + { 2 M a^2 \over r_{+} } } ,
 \label{eq:Risco_kerr}
\end{equation}
where the so-called Kerr parameter $a = J/M=Mj$ is the angular momentum per unit mass.

%%%%%%%%%%%%%%%%%%%%%%%%%%%%%
\section{Numerical results}
\label{sec:results}
%%%%%%%%%%%%%%%%%%%%%%%%%%%%

%%%%%%%%%%%%%%%%
\subsection{EOS sensitive relations}
\label{sec:eos_sen_rel} 
%%%%%%%%%%%%%

\begin{figure}
  \centering
  \includegraphics*[width=9.0cm]{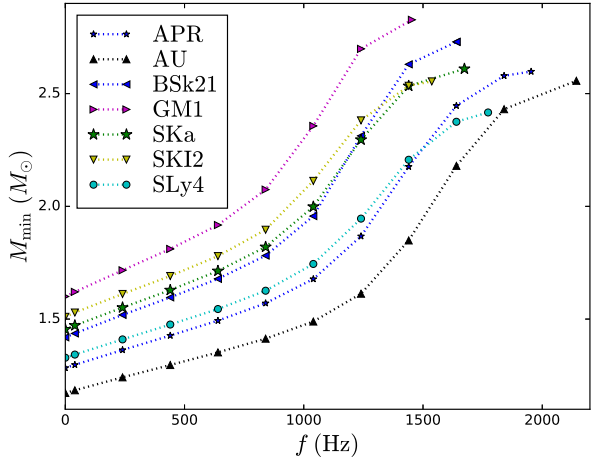}
  \caption{Minimum required mass $M_{\rm min}$ for the appearance of ISCO is plotted against the neutron star spin frequency $f$ for some of our EOS models.}
  \label{fig:Min_M_f}
\end{figure}

Before presenting our universal relations for the radius and orbital frequency of ISCO, we first study and review how the ISCO generally depends on the EOS and stellar spin frequency 
\citep{Miller:1998_793,Bhattacharyya:2011_3247,Torok:2014_L5}. 
Let us first consider the situation for nonrotating neutron stars where 
the ISCO radius is given analytically by $R_{\rm ISCO} = 6 M$. We must require the stellar radius $R < R_{\rm ISCO}$ in order for the ISCO to exist outside the star. The compactness of a 
star needs to be $M/R > 1/6$ in order for the ISCO to exist. Nonrotating neutron stars constructed 
from typical EOS models can easily achieve this condition when $M$ is high enough.  
The situation for rotating stars is more complicated since one also needs to consider the stellar spin frequency as a parameter. In general, for a given EOS model, there exists a minimum mass $M_{\rm min}$ at each value of the stellar spin frequency $f$ in order for the ISCO to exist (i.e., $R < R_{\rm ISCO}$). In Figure~\ref{fig:Min_M_f}, we plot $M_{\rm min}$ against $f$ to illustrate this fact using some of our EOS models. 
It is seen that $M_{\rm min}$ rises with $f$ generally and it also depends sensitively on the EOS. In particular, $M_{\rm min}$ can change by as much as 100\% for different EOSs at high values
of $f$.

While the value of $M_{\rm min}$ depends sensitively on $f$ and the EOS, Figure~\ref{fig:Min_M_f} shows that an ISCO can exist generally only for high mass neutron stars. 
For rapidly rotating models with $f > 1500$ Hz, an ISCO exists only for $M_{\rm min} > 2 M_\odot$ for the EOS models plotted in Figure~\ref{fig:Min_M_f}. 
Nevertheless, it should be noted that the currently observed fastest rotating neutron star has a 
frequency of about 700 Hz \citep{Hessels:2006_1901}. The corresponding smallest possible value 
of $M_{\rm min}$ for the existence of ISCO at that frequency is about $1.4 M_\odot$ as shown in 
Figure~\ref{fig:Min_M_f}.

\begin{figure}
  \centering
  \includegraphics*[width=9.0cm]{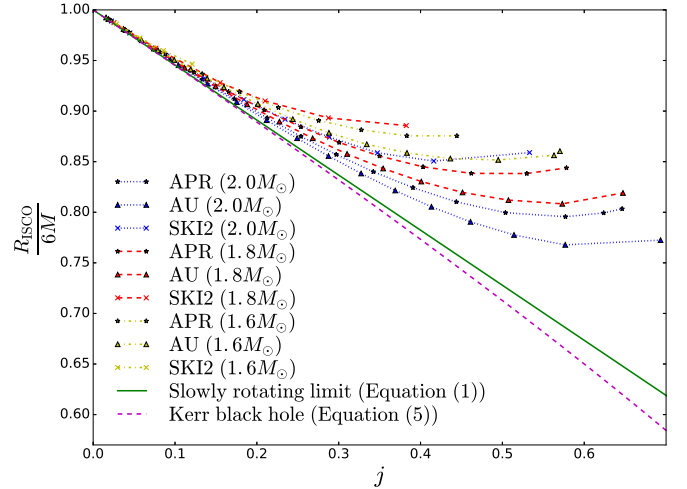}
  \caption{ISCO radius $R_{\rm ISCO}$ (normalized by $6M$) is plotted against the dimensionless
  spin parameter $j$ for different masses and EOS models. The results for the slowly rotating 
  limit (Equation~(\ref{eq:Risco_slow})) and the Kerr black hole (Equation~(\ref{eq:Risco_kerr}))
  are also shown.   }
  \label{fig:Risco_j}
\end{figure}

Let us now study the dependence of the ISCO radius on the EOS and stellar properties. Figure~\ref{fig:Risco_j} plots $R_{\rm ISCO}$ (normalized by $6M$) against the spin parameter $j$ 
for three EOS models. We also consider three different masses 1.6, 1.8, and 2.0 $M_\odot$ for each 
EOS. In the figure, we also plot the results given by the slowly rotating limit and the Kerr black hole for comparison. 
It is seen that the numerical results for rotating stars agree quite well to the prediction of 
Equation~(\ref{eq:Risco_slow}) for slow spin cases $j \lesssim 0.1$. The numerical results also
converge to the nonrotating limit $R_{\rm ISCO} / 6 M = 1$ as $j$ decreases to zero.  
However, the results deviate significantly from the prediction of the slowly rotating limit for high values of $j$. It is also apparent that $R_{\rm ISCO}$ becomes very sensitive to the EOS 
and the stellar mass as the value of $j$ increases. 

It can also be seen from Figure~\ref{fig:Risco_j} that $R_{\rm ISCO}$ for the Kerr black hole (dashed line) behaves quite differently compared to rapidly rotating neutron stars. In particular, for a given value of $j$, the ISCO radius of a rapidly rotating neutron star (black hole) is larger (smaller) than that predicted by the slowly rotating limit (solid line). This difference may not be surprising since it is well known that the exterior spacetime of rapidly rotating neutron stars cannot be approximated by the Kerr metric.

The contributions from higher-order multipole moments of the rotating star become significant for high values of $j$.
Figure~\ref{fig:Risco_j} illustrates that the effects of higher-order multipole moments, presumably dominated by the quadrupole moment $Q$, generally increase the ISCO radius for rotating
neutron stars. 
The properties of the ISCO in terms of the multipole moments of the rotating neutron star and the spacetime have been studied \citep{Shibata:1998_104011,Berti:2004_1416,Berti:2005_923}.
In particular, the dependence of $R_{\rm ISCO}$ on $j$ and $Q$ has been studied by \citet{Pappas:2015_4066}. We shall discuss the connection of this work to our proposed universal relations in Section~\ref{sec:other_relations}.

%%%%%%%%%%%%%%%%%%%%%%%%%%
\subsection{Universal relations}
\label{sec:uni_relation}
%%%%%%%%%%%%%%%%%%%%%%%%%%%

Having seen that the ISCO radius $R_{\rm ISCO}$ for rapidly rotating neutron stars depends sensitively on the EOS and stellar properties such as $M$ and $f$, it is thus somewhat surprising that approximate EOS-insensitive relations connecting the ISCO to the global stellar quantities 
can indeed be found. To motivate our proposed universal relations, let us first note that previously known universal relations, as mentioned in Section~\ref{sec:intro}, are relations connecting suitable dimensionless quantities.

\begin{figure}
  \centering
  \includegraphics*[width=9.0cm]{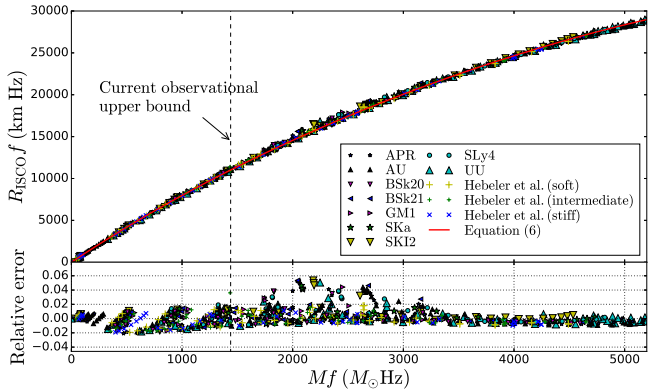}
  \caption{Universal relation connecting the ISCO radius $R_{\rm ISCO}$ to the neutron star spin   frequency $f$ and mass $M$. The lower panel shows the relative error between the numerical data and the fitting curve given by Equation~(\ref{eq:Risco_fit}). }
  \label{fig:Risco_uni}
\end{figure}

In order to search for possible universal relations concerning the ISCO, we construct dimensionless quantities out of the relevant physical variables such as $R_{\rm ISCO}$, $M$, and $f$ in this situation. In particular, we consider the 
two quantities $R_{\rm ISCO} f$ and $M f$, which are dimensionless in geometric units.
In Figure~\ref{fig:Risco_uni}, we plot $R_{\rm ISCO} f$ against $M f$ using our 12 different EOS models. In contrast to Figure~\ref{fig:Risco_j}, we now see that the results are relatively 
EOS-insensitive for a large range of $Mf$. To facilitate any potential use of this universal relation in astrophysical situations, we have restored physical units in the figure so that $R_{\rm ISCO}$, $M$, and $f$ are expressed in km, $M_\odot$, and Hz, respectively. 
The lower panel of Figure~\ref{fig:Risco_uni} shows the relative error, $({\hat y} - y)/y$, between the numerical data $\hat y$ and a fitting curve (solid line) $y$ given by 
\begin{equation} 
y = a_1 x + a_2 x^2 + a_3 x^3 + a_4 x^4  ,
\label{eq:Risco_fit}
\end{equation} 
where $y = R_{\rm ISCO} f$ and $x = M f$. 
The fitting parameters are $a_1 = 8.809$, $a_2 = -9.166 \times 10^{-4}$, $a_3 = 8.787 \times 10^{-8}$, and $a_4 = - 6.019 \times 10^{-12}$. 
The vertical dashed line in the figure corresponds to the current observational upper bound of
$Mf$ in the sense that its value is obtained by combining the maximum measured mass at about $2 M_\odot$ \citep{Demorest:2010_1081,Antoniadis:2013_6131} and the largest spin frequency $f = 716$ Hz for the fastest rotating neutron star PSR J1748-2446ad \citep{Hessels:2006_1901}. We see that the numerical data can generally be fitted by Equation~(\ref{eq:Risco_fit}) to about 2\% for a large range of $Mf$, except for a few outsiders between $Mf\approx 2000$ and 3000 $M_\odot {\rm Hz}$ 
that have relative errors close to 6\%.

As we shall see in Section~\ref{sec:other_relations}, to second order in the spin parameter $j$, the ISCO radius $R_{\rm ISCO}$ can be determined by $M$, $j$, and the quadrupole moment $Q$ of the rotating star (see Equation~(\ref{eq:Risco_shibata})). While $M$ is one of the common observables for neutron stars, the other two quantities $j$ and $Q$ cannot be obtained easily from observations. By comparison, our approximate EOS-insensitive relation (Equation~(\ref{eq:Risco_fit})) is more appealing and would be more useful since the relation is also valid for rapidly rotating stars and the relevant quantities $M$ and $f$ are more promising observables for neutron stars. In particular, these quantities can usually be measured to high accuracy if they are observed.

\begin{figure}
  \centering
  \includegraphics*[width=9.0cm]{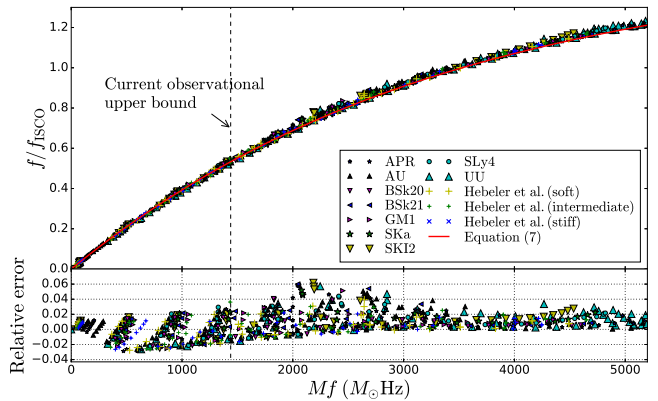}
  \caption{Universal relation connecting the ISCO frequency $f_{\rm ISCO}$ to the neutron star
  spin frequency $f$ and mass $M$. The lower panel shows the relative error between the numerical 
  data and the fitting curve given by Equation~(\ref{eq:Fisco_fit}). }
  \label{fig:Fisco_uni}
\end{figure}

Having seen that there exists a universal relation connecting $R_{\rm ISCO}$ to the global 
stellar quantities $M$ and $f$, it is thus not surprising that a similar universal relation should
also exist for the orbital frequency $f_{\rm ISCO}$ of the ISCO, since $f_{\rm ISCO}$ is determined
by the exterior spacetime metric and the position of ISCO around a rotating star. 
As we shall discuss in Section~\ref{sec:discuss}, $f_{\rm ISCO}$ is indeed more interesting than $R_{\rm ISCO}$ since it may be directly related to the observed QPOs in LMXBs.  
In Figure~\ref{fig:Fisco_uni}, we plot the ratio $f/f_{\rm ISCO}$ against $Mf$ to demonstrate the
universality for $f_{\rm ISCO}$. As before, we have restored physical units so that $M$ is 
expressed in $M_\odot$ and the frequencies $f$ and $f_{\rm ISCO}$ are expressed in Hz. The vertical dashed line corresponds the current observational upper bound of $Mf$. The solid line is a fitting curve given by 
\begin{equation}
y = b_1 x + b_2 x^2 + b_3 x^3 + b_4 x^4 , 
\label{eq:Fisco_fit}
\end{equation}
where $y = f / f_{\rm ISCO}$ and $x = Mf$.  
The fitting parameters are given by $b_1 = 4.497 \times 10^{-4}$, $b_2 = -6.130 \times 10^{-8}$, $b_3 = 4.527 \times 10^{-12}$, and $b_4 = - 1.446 \times 10^{-16}$. The relative error between the numerical data and the fitting curve is plotted in the lower panel of Figure~\ref{fig:Fisco_uni}. Similar to Figure~\ref{fig:Risco_uni}, we see that the relation is approximately EOS-insensitve to 
a few percent level in a large range of $Mf$.

%%%%%%%%%%%%%%%%%%%%%%%
\section{Analysis}
\label{sec:analy}
%%%%%%%%%%%%%%%%%%%%%%%
  
%%%%%%%%%%%%%%%%%%%%%%%
\subsection{Connection to Kerr black hole}
\label{sec:kerr_BH}
%%%%%%%%%%%%%%%%%%%%%%%%%%

As we have seen in Figure~\ref{fig:Risco_j}, the dependence of $R_{\rm ISCO}$ on $M$ and $j$ for rotating neutron stars is very different from that for the Kerr black hole. 
The difference is due to the fact that the multipolar structure of the spacetime of a rapidly rotating star \citep{Laarakkers:1999_282,Berti:2004_1416,Pappas:2012_231104,Yagi:2014_124013} is in general different from that of a (uncharged) rotating black hole, which depends only on two parameters, $M$ and $j$, because of the no-hair theorem. 
It should be noted that approximate analytic formulas for $R_{\rm ISCO}$ and $f_{\rm ISCO}$ around 
a rotating star in terms of the various moments of the star have been obtained by \citet{Shibata:1998_104011}. 
For comparison, the approximately EOS-insensitive relations that we find for $R_{\rm ISCO}$ (Equation~(\ref{eq:Risco_fit})) and $f_{\rm ISCO}$ (Equation~(\ref{eq:Fisco_fit})) depend 
only on two parameters, $M$ and $f$. It is thus interesting to see whether these universal relations for rotating neutron stars share greater similarities to the corresponding relations for the Kerr black hole or not. 

\begin{figure}
  \centering
  \includegraphics*[width=9.0cm]{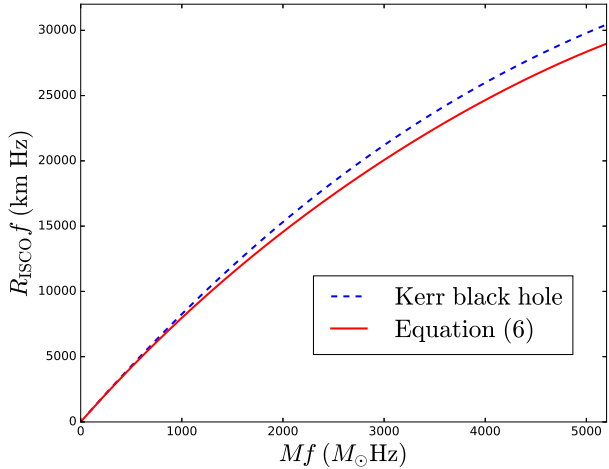}
  \caption{Comparison between the universal relation for $R_{\rm ISCO}$ for neutron stars
  (Equation~(\ref{eq:Risco_fit})) and the corresponding relation for the Kerr black hole.}
  \label{fig:Risco_uni_Kerr}
\end{figure}

The ISCO radius $R_{\rm ISCO}$ around a Kerr black hole is determined analytically by 
Equation~(\ref{eq:Risco_kerr}) in terms of the mass $M$ and spin parameter $j$ of the black hole. 
Since our universal relation (Equation~(\ref{eq:Risco_fit})) for $R_{\rm ISCO}$ around rotating stars involves the spin frequency $f$ instead of $j$, we replace $j$ by the spin frequency 
$f_{\rm H} = \Omega_{\rm H} / 2 \pi$ of the black hole horizon \citep[e.g.,][]{Wald:book}
in order to obtain the corresponding equation for a rotating black hole. The angular frequency $\Omega_{\rm H}$ of the black hole horizon can be expressed in terms of $M$ and $j$:
\begin{equation}
2 M \Omega_{\rm H} = { j \over 1 + \sqrt{1 - j^2} } . 
\label{eq:Omega_H}
\end{equation}  
Solving Equation~(\ref{eq:Omega_H}) for $j$ and using the result in Equation~(\ref{eq:Risco_kerr}) 
then provide us with the desired relation connecting $R_{\rm ISCO}$, $M$, and $\Omega_{\rm H}$ (and 
hence $f_{\rm H}$). 
In Figure~\ref{fig:Risco_uni_Kerr}, we compare the universal relation for $R_{\rm ISCO}$ (Equation~(\ref{eq:Risco_fit})) for rotating neutron stars and the corresponding relation for the 
Kerr black hole. We see that the two relations agree quite well over a large range of 
$Mf$. The relative difference between the two relations is within 6\% up to $Mf = 5000 M_\odot {\rm Hz}$, which is well above the current observational upper bound for rotating 
neutron stars.

In contrast to the plot of $R_{\rm ISCO}$ against $j$ in Figure~\ref{fig:Risco_j}, we now see that the relations between the dimensionless quantities $R_{\rm ISCO}f$ and $Mf$ for neutron stars and the Kerr black hole are very close to each other for the range of $Mf$ allowed theoretically
for neutron stars.  
This similarity is quite intriguing if one recalls that the multipolar structure of the spacetime
of a rapidly rotating neutron star is generally different from that of the Kerr black hole. 
Furthermore, the spin frequency $f_{\rm H}$ of the black hole horizon is conventionally defined 
as the orbital frequency of a zero-angular-momentum observer at the horizon due to the effect of
frame dragging \citep{Wald:book}. The concept of $f_{\rm H}$ is thus also quite different from the spin frequency of a rotating star, which is associated with the fluid motion of the star itself and has a clear Newtonian limit.

%%%%%%%%%%%%%%%%%%%%%%%
\subsection{Connection to other universal relations} 
\label{sec:other_relations}
%%%%%%%%%%%%%%%%%%%%%%%%%%%%%%%%

While there is no exact analytic solution for the spacetime of rapidly rotating compact stars
in general relativity, it was shown by \citet{Pappas:2013_3007} that the two-soliton analytic solution of \citet{Manko:1995_3063} can provide an accurate representation of the exterior spacetime of rotating neutron stars. 
By further making use of the EOS-insensitive ($\sim 10$\% variability) universal relations between the multipole moments of rapidly rotating neutron stars \citep{Pappas:2014_121101,Yagi:2014_124013}, the formally four-parameter two-soliton solution can lead to an EOS-independent description of the spacetime in terms of the mass $M$, spin parameter $j$, and the reduced quadrupole moment $\alpha\equiv - Q/(j^2 M^3)$ of a rotating neutron star \citep{Pappas:2015_4066}. In retrospect, it may thus not be entirely surprising that there can exist universal relations for the ISCO radius and frequency, although the fact that these relations (Equations~(\ref{eq:Risco_fit})-(\ref{eq:Fisco_fit})) have relatively simple forms is very interesting and unexpected. 

In \citet{Pappas:2015_4066}, the properties of ISCO are studied using the above mentioned analytic
spacetime. In particular, the region of the parameter space $(j, \alpha)$ for which the ISCO is outside the stellar surface ($R_{\rm ISCO} > R$) is investigated. It is found that $\alpha$ is 
restricted to be smaller than about 4.5 for small spins at $j \approx 0.1$ and it is 
further restricted to $\alpha \lesssim 2$ at $j \approx 0.7$ (see Figure 3 of \citet{Pappas:2015_4066}). For comparison, we note that $\alpha=1$ for the Kerr black hole. 
As we shall discuss later, the restriction on the value of $\alpha$ might be responsible for the 
similarity between Equation~(\ref{eq:Risco_fit}) and the corresponding relation for the Kerr 
black hole.

A formula for the ISCO radius around rotating relativistic stars as an expansion in the mulitpole moments of the spacetime has been developed by \citet{Shibata:1998_104011}. 
\citet{Pappas:2015_4066} found that, for the region of the parameter space $(j, \alpha)$ where $R_{\rm ISCO} > R$, the \cite{Shibata:1998_104011} formula agrees to the prediction of the two-soliton solution to within 1\%. In the following, we shall make use of the expansion formula and other universal relations for rotating neutron stars to provide a connection to our universal relation for $R_{\rm ISCO}$. 

The \citet{Shibata:1998_104011} expansion formula for $R_{\rm ISCO}$, up to the third order, is 
given by 
\begin{eqnarray}
R_{\rm ISCO} &=& 6 M \left( 1 - 0.54433 j - 0.22619 j^2 + 0.17989 \alpha j^2 \right. \nonumber \\
\cr
&& \left. - 0.23002 j^3  + 0.26296 \alpha j^3 - 0.05317 j^3 \beta \right) , 
\label{eq:Risco_shibata}
\end{eqnarray} 
where $\beta \equiv - J_3 /(j^3 M^4)$ and $J_3$ is the spin octupole moment. As discussed by 
\citet{Pappas:2015_4066}, the parameter $\beta$ can be given in terms of $\alpha$ using the 
universal relation between the quadrupole and spin octupole moments \citep{Pappas:2014_121101}:
\begin{equation}
\beta^{1/3} = -0.36 + 1.48 \alpha^{0.65/2} .
\label{eq:Q_J3}
\end{equation}   
Equation~(\ref{eq:Risco_shibata}) provides us with a starting point to construct a universal relation for the ISCO radius. For the Kerr black hole, we use Equation~(\ref{eq:Omega_H}) to replace $j$ by $Mf$ in order to obtain the corresponding universal relation for the ISCO radius. For rotating neutron stars, we can make use of the relation connecting $Mf$, $j$, and $\alpha$ found numerically by \citet{Pappas:2015_4066}:
\begin{eqnarray}
{ {Mf} \over j} &=& B_0 + B_1 j + B_2 j^2 + \left( A_0 + A_1 j + A_2 j^2 \right) 
\left( \sqrt{\alpha} \right)^{n_1}  \nonumber \\
\cr
&& + \left( C_0 + C_1 j + C_2 j^2 \right) \left( \sqrt{\alpha} \right)^{n_2} ,
\label{eq:Pappas_Mfj}
\end{eqnarray} 
where the fitting parameters are $A_0=15.0297$, $A_1=-0.114154$, $A_2=-7.72439$, $B_0=-1.48338$,
$B_1=-1.07874$, $B_2=1.64592$, $C_0=-2.45303$, $C_1=3.40995$, $C_2=7.39354$, $n_1=-1.1698$, 
and $n_2 = -4.50216$. This relation is EOS-insensitive to within the 1\% level.

In order to use Equations~(\ref{eq:Risco_shibata})$-$(\ref{eq:Pappas_Mfj}) to construct a relation 
between $R_{\rm ISCO}f$ and $Mf$, we still need one more relation to connect $\alpha$ and $j$. 
As discussed above, \citet{Pappas:2015_4066} studied the properties of ISCO based on the two-soliton
analytic solution and in particular he considered the contours of constant relative difference
between $R_{\rm ISCO}$ and $R$, defined by the parameter 
$\delta R = (R_{\rm ISCO} - R)/R_{\rm ISCO}$, in the parameter space of $(j, \alpha)$. Note that the region with $\delta R \geq 0$ corresponds to the parameters where the ISCO can exist outside the
stellar surface. 
We refer the reader to Figure 3 of \citet{Pappas:2015_4066} to see how different values of $\delta R$ define different EOS-independent contours in the parameter space.
Those contours with $\delta R \geq 0$ thus provide us with the desired relations between $\alpha$ and $j$ connected to the ISCO. We consider the contours defined by $\delta R = 0$, 0.11, 0.22, and 0.33 in Figure 3 of \citet{Pappas:2015_4066} and use them to express $\alpha$ in terms of $j$ numerically in Equation~(\ref{eq:Pappas_Mfj}), which is then solved together with Equations~(\ref{eq:Risco_shibata}) and (\ref{eq:Q_J3}) in order to construct a relation between $R_{\rm ISCO}f$ and $Mf$.

\begin{figure}
  \centering
  \includegraphics*[width=9.0cm]{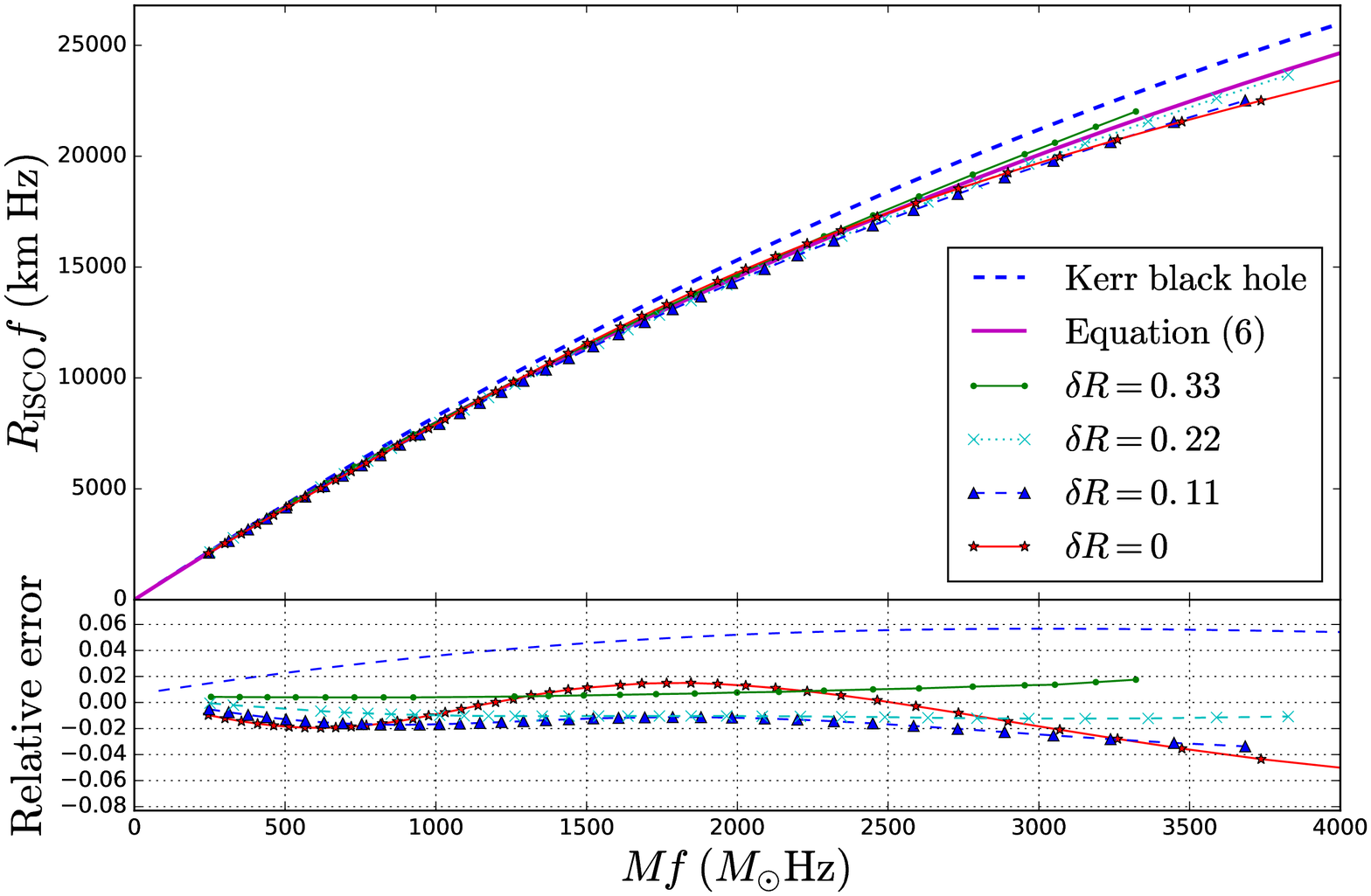}
  \caption{Universal relations constructed from 
  Equations~(\ref{eq:Risco_shibata})$-$(\ref{eq:Pappas_Mfj}) for different constant 
  $\delta R$ contours in the parameter space $(j,\alpha)$ extracted from Figure 3 of 
  \citet{Pappas:2015_4066}. Equation~(\ref{eq:Risco_fit}) and the corresponding relation for the Kerr black hole are also plotted for comparison. The lower panel shows the relative errors, taking 
  Equation~(\ref{eq:Risco_fit}) as a reference.   }
  \label{fig:Risco_uni_pappas}
\end{figure}

It should be pointed out that different contours of constant $\delta R$ lead to different relations between $Mf$ and $j$ using Equation~(\ref{eq:Pappas_Mfj}). There is thus no a priori reason to guarantee that our procedure would lead to a single universal relation connecting $R_{\rm ISCO}f$
and $Mf$.
In the upper panel of Figure~\ref{fig:Risco_uni_pappas}, we plot the resulting relations between $R_{\rm ISCO}f$ and $Mf$ obtained for the different contours. 
Equation~(\ref{eq:Risco_fit}) and the corresponding relation for the Kerr black hole are also 
plotted in the figure for comparison. The relative errors, taking Equation~(\ref{eq:Risco_fit}) as a reference, are shown in the lower panel of Figure~\ref{fig:Risco_uni_pappas}. 
It is interesting to note that the relations corresponding to different contours of $\delta R$ turn out to match each other very well and establish approximately the existence of a single universal relation. 
Deviations between the different $\delta R$ curves start to become noticeable at the high end of $Mf$. These $\delta R$ curves also match Equation~(\ref{eq:Risco_fit}) very well to within about
2\% for a large range of $Mf$, which is consistent with the accuracy level of Equation~(\ref{eq:Risco_fit}) as a fitting curve to the numerical data.

In Figure~\ref{fig:Risco_uni_pappas}, it can be seen that the different $\delta R$ curves generally tend toward the Kerr case as the value of $\delta R$ increases. 
This is consistent with the fact that the value of $\alpha$, at a fixed value of $j$, on a contour of constant $\delta R$ decreases as the value of $\delta R$ increases as shown in Figure 3 of 
\citet{Pappas:2015_4066}. In particular, the maximum value of $\alpha$ on a contour decreases from about 4.5 to 2 as $\delta R$ increases from 0 to 0.33, and thus the value of $\alpha$ indeed tends toward the Kerr case $\alpha = 1$. 
The restriction on the value of $\alpha$ for the existence of an ISCO might be responsible for the
similarity between Equation~(\ref{eq:Risco_fit}) and the corresponding relation for the Kerr black 
hole. However, further work is needed to study this connection in more detail.

%%%%%%%%%%%%%%%%%%%%%%%%
\section{Discussion}
\label{sec:discuss}
%%%%%%%%%%%%%%%%%%%%%%%%%%

As mentioned in Section~\ref{sec:eos_sen_rel}, an ISCO can exist only around high mass neutron stars. Depending on the underlying EOS model, the minimum mass for the existence of an ISCO can increase from about 1.4 to $2 M_\odot$ as the spin frequency of the star increases from 700 Hz to 1500 Hz. 
In this paper, we have proposed two universal relations to connect the ISCO radius and orbital 
frequency (when an ISCO exists) to the spin frequency and mass of rotating neutron stars. We have
also compared the universal relation for the ISCO radius to the corresponding relation for the Kerr black hole and analyze it from the perspective of other known universal relations for rotating neutron stars. In the following, we shall discuss how the ISCO universal relations can be applied practically.  

As mentioned briefly in Section~\ref{sec:intro}, the ISCO may be relevant to the kHz QPOs well observed from neutron stars in LMXBs. These QPOs often come in pairs, one with a lower frequency $f_{\rm l}$ and another with a higher frequency $f_{\rm u}$. Many models have been proposed to explain the kHz QPOs. We refer the reader to \citet{Klis:2006_39}, \citet{Torok:2016_273}, and
references therein for reviews.

While there is still no general consensus on the mechanism for generating the kHz QPOs, it is believed that the QPOs are associated with the orbital motion and/or oscillations near the inner edge of an accreting disk surrounding the central neutron star. 
The ISCO defines the smallest inner radius that an accreting disk around a neutron star can have. Indeed, there is also evidence to suggest that the ISCO is responsible for some of the 
observed properties associated to the kHz QPOs. For instance, it has been observed 
that the quality factor of the kHz QPOs drops sharply in some systems, which is consistent with
the signature that the orbits involved approach the ISCO \citep{Barret:2006_1140,Barret:2007_1139, Boutelier:2009_1901}. Our proposed universal relations for the ISCO are relevant and applicable to these systems. 

As an illustration of potential applications of our proposed universal relations, we apply Equation~(\ref{eq:Fisco_fit}) to the system 4U 0614+09. The spin frequency $f=414$ Hz of the neutron star in the system has been inferred from the detection of burst oscillations during a thermonuclear X-ray burst \citep{Strohmayer:2008_L37}. The highest frequency for the upper kHz QPO is measured to be $f_{\rm u} \approx 1220$ Hz \citep{Boutelier:2009_1901}. Using these two observed frequencies $f$ and $f_{\rm u}$, together with the assumption that $f_{\rm u}$ is equal to the ISCO orbital frequency $f_{\rm ISCO}$, we apply Equation~(\ref{eq:Fisco_fit}) to determine that the mass of the neutron star is $2.0 M_\odot$. We can then determine the ISCO radius to be about 16 km by using Equation~(\ref{eq:Risco_fit}), and hence put a constraint on the radius of the star to be 
$R < 16$ km.

If, in reality, the QPO frequency $f_{\rm u}$ corresponds to the orbital frequency at an orbit outside the ISCO, instead of being equal to $f_{\rm ISCO}$, then our value $2.0 M_\odot$ would 
set an upper bound for the mass of the star. 
Previously, \citet{Boutelier:2009_1901} also estimated the mass of 4U 0614+09 to be about $1.9 M_\odot$ by using the value $f_{\rm ISCO}=1250$ Hz, which is obtained by using the observed drop of the quality factor of the lower kHz QPO and extrapolating it to zero, together with the constant
frequency difference between the lower and upper kHz QPOs. 
In their analysis, \citet{Boutelier:2009_1901} determined the mass by assuming slow rotation and included the first-order correction due to the dimensionless spin parameter $j$, the value of which generally depends on the mass and EOS for a given spin frequency of the star. For comparison, besides the minimal assumption that $f_{\rm u} = f_{\rm ISCO}$, our analysis applies generally to rapidly rotating stars and is insensitive to the underlying EOS model to within about the 2\% level. 

If the existence of an ISCO around some neutron stars in LMXBs can be confirmed, together with 
the measurement of the ISCO frequency (or radius), our universal relations can be used to yield a 
precise determination of the masses of the neutron stars, bypassing the assumption of slow
rotation and the uncertainty related to the dimensionless spin parameter, which are commonly
required in the literature.

%\section*{Acknowledgments}
We thank the anonymous referee for very helpful comments that improved the quality of the manuscript. In particular, we thank the referee for bringing the work of \citet{Pappas:2015_4066} to our attention. Some of our chosen EOSs are obtained from the EOS repository CompOSE (compose.obspm.fr). We thank the team members of CompOSE and LORENE for their development.

%%%%%%%%%%%%%%%%%
%% References
%%%%%%%%%%%%%%%%%

%% \bibliography{biblio}

%%%%%%%%%%%%%%%%

\end{document}